# Optimized Dynamical Decoupling in Ξ-Type *n*-Level Quantum Systems

LINPING CHAN and SHUANG CONG[*]

Department of Automation
University of Science and Technology of China
Jinzhai Road 96, Hefei, Anhui, 230027
P. R. CHINA
[*] Corresponding author's email:scong@ustc.edu.cn

*Abstract:* - In this paper, we first design a type of Bang-Bang (BB) operation group to reduce the phase decoherence in a Ξ-type *n*-level quantum system based on the dynamical decoupling mechanism. Then, we derive two kinds of dynamical decoupling schemes: periodic dynamical decoupling (PDD) and Uhrig dynamical decoupling (UDD). We select the non-diagonal element of density matrix as a reference index, and investigate the behavior of quantum coherence of the Ξ-type *n*-level atom under these two dynamical decoupling schemes proposed. At last, we choose a Ξ-type six-level atom as a system controlled, and use the decoupling schemes proposed to suppress the phase decoherence. The simulation experiments and the comparison results are given.

## 1 Introduction

In the quantum world, the coherence and entanglement are the sources of power that make quantum computation [1] surpass classical computation. But quantum systems always interact with their surrounding environment realistically to some extent. No matter how weak the coupling is, the evolution of the quantum system is eventually plagued by non-unitary features like decoherence and dissipation [2]. Decoherence destroys the coherence of the quantum superposition states in the process of evolution, and results in the reduction or even erosion of the entanglement between subsystems. The design of strategies being able to protect the evolution of a quantum system against decoherence represents a challenging conceptual issue. In recent years, many strategies have been proposed to counteract the effects of environmental couplings successfully in open-system evolutions. Dynamical decoupling [3] (known as quantum BB control) is one of the universal methods originally introduced in order to overcome decoherence-errors. It makes use of the coherent averaging effects [4] and uses the twin-born tailored powerful pulses to average out the effect of unwanted Hamiltonian. But up to now the dynamical decoupling pulse sequences are almost designed based on dividing the total evolution time into equidistant time periods, it has been further shown that in principle it is possible to use optimized pulse sequences [5]-[6]. In [5], the UDD was discussed in the spin-boson model. And the potential of non-equidistant pulse sequences was demonstrated by concatenated pulse sequences in [6]. But the high-dimensional systems have attracted much interest for their applications in quantum control and quantum computation [7]. So the related researches in high-dimensional systems are needed to be studied.

This paper will study the effectiveness of the optimized dynamical decoupling scheme UDD in arbitrary *n*-level atom in Ξ-configuration. The key point is to put on the comparison of the effectiveness of suppressing the decoherence between the standard periodic dynamical decoupling scheme (PDD) [3] and UDD. Toward this goal, we first study the dynamics of a decoherence Ξ-type *n*-level atom driven by dynamical decoupling, and design the corresponding BB operation group of the dynamical decoupling scheme. Then according to different time intervals between the adjacent BB pulses, two decoupling schemes: PDD and UDD are designed, and then select the non-diagonal element of density matrix [3], [8] as a reference index to investigate the behavior of quantum coherence of the Ξ-type *n*-level atom under dynamical decoupling schemes. At last, a Ξ-type six-level atom is taken as an example. The numerical simulation is carried out and the analysis is given.

The paper is arranged as follows. In Section II, a brief derivation of the dynamical decoupling mechanism in arbitrary *n*-level atom in Ξ-configuration is presented, and the corresponding BB operation group is designed, then PDD and UDD are introduced. In Section III, the behavior of quantum coherence of the Ξ-type *n*-level atom under two dynamical decoupling schemes is investigated, then the numerical simulation of a Ξ-type six-level atom is carried out, and a comparison of the results is analyzed. In Section IV, we give a brief summary and close by discussing possibilities for future work.

## 2 Dynamical Decoupling Mechanism

The core ideology of dynamical decoupling which is termed as "quantum BB control" is to eliminate the interaction Hamiltonian with the bath from the total Hamiltonian for the whole system. It uses the tailored unitary pulses, the impulsive full-power operations, and can be turned on/off for negligible amount of time with ideally arbitrarily large strength to average out the interaction Hamiltonian between the quantum system and the bath which results in decoherence. Each pulse in the pulse sequence represents a unitary BB operator, and the pulse sequence represents a BB operation group. Thus, the key point to design the BB control field is to design the BB operation group, and the operators in BB operation group should be the least. So in this section, we first design the BB operation group under the phase decoherence in an arbitrary *n*-level atom with Ξ-configuration, then according to different time intervals between the adjacent BB pulses, we derive two decoupling schemes: PDD and UDD.

### 2.1 The Design of BB Decoupling Operators

Consider an arbitrary *n*-level atom in Ξ-configuration, we assume that the dynamics between the bath and the system can be described by the following interaction Hamiltonian [8]:

$$H_{SB} = \hbar \sum_{i=0}^{n-2} \sum_{ki} (c_{ki}\sigma_z^{(i+1,i)} + b_{ki}\sigma_x^{(i+1,i)})(j_{ki}a_{ki}^+ + j_{ki}^* a_{ki}) \quad (1)$$

where $c_{ki}$ and $b_{ki}$ are the coefficients of the relative magnitude of the phase decoherence and the amplitude decoherence, respectively, and $\{j_{ki}\}$ are the coupling constants for virtual exchanges of excitations with the thermal reservoirs. According to the changes of $c_{ki}$ and $b_{ki}$, we can conclude from (1) that:

1) if $c_{ki} = 0, b_{ki} \neq 0$, then the reservoir is an adiabatic reservoir which results in the amplitude damping;

2) if $c_{ki} \neq 0, b_{ki} = 0$, then the reservoir is a thermal reservoir which results in the phase damping;

3) if $c_{ki} \neq 0, b_{ki} \neq 0$, then the reservoir is a general reservoir which results in the general decoherence.

Suppose energy exchange processes typically involve time scales much longer than decoherence mechanisms, we neglect the effects associated to quantum amplitude damping [9]. In such a case, the interaction Hamiltonian can be got if we choose $c_{ki} = 1, b_{ki} = 0$ for simplicity, and it can be written as

$$H_{SB} = \hbar \sum_{i=0}^{n-2} \sum_{ki} \sigma_z^{(i+1,i)} (j_{ki}a_{ki}^+ + j_{ki}^* a_{ki}) \quad (2)$$

Assume that the BB operation group is expressed as

$$G = \{g_m\} \, (m = 1, 2, \cdots, |G|-1),$$

where $|G|$ represents the number of the operations, and $g_m$ is a constant pulse matrix.

Suppose a quantum system has the evolution time $T$, and divide $T$ into $N$ cycles $T_c$, according to the core ideology of dynamical decoupling we apply the BB operation group $G = \{g_m\}$ to eliminate the unwanted Hamiltonian $H_{SB}$ in (2) in every decoupling cycle. And the dynamic decoupling condition can be expressed as

$$\frac{1}{|G|} \sum_{m=0}^{|G|-1} g_m^+ H_{SB} g_m = 0 \quad (3)$$

In [10], the BB operation group to suppress the phase decoherence in an arbitrary *n*-level atom with

Ξ-configuration was designed. And it can be written as

$$G = \{I, g_1, g_2, \cdots, g_{n-1}\}$$

in which $|G| = n$; $I$ represents a unit matrix with order $n$; $g_1, g_2, \cdots, g_{n-1}$ can be expressed as

$$g_1 = \exp(i * \sigma_x^{(1,0)} * \pi / 2) * \cdots * \exp(i * \sigma_x^{(k+1,k)} * \pi / 2)$$
$$* \exp(i * \sigma_x^{(k+2,k+1)} * \pi / 2) * \cdots * \exp(i * \sigma_x^{(n-1,n-2)} * \pi / 2)$$
(4a)

$$g_2 = (g_1)^2, \cdots, g_{n-1} = (g_1)^{n-1} \quad (4b)$$

From the above analysis, one can see that there are $N*|G|$ BB operators (pluses) in the evolution time $T$. Let $t_i$ ($i = 1, 2, \cdots, N*|G|$) represent the moments when the BB operators are produced, so except that there is a operator at $t_{N*|G|} = T$, in the time interval $(0, T)$ the number of the pluses $M$ is

$$M = |G| * N - 1. \quad (5)$$

According to different choices of $t_i$, one can obtain many kinds of dynamical decoupling schemes. Next we'll introduce the Periodic dynamical Decoupling (PDD) and Uhrig Dynamical Decoupling (UDD).

### 2.2 Periodic Dynamical Decoupling (PDD)

Periodic dynamical decoupling (PDD) is the basic dynamical decoupling scheme. It uses the twin-born tailored powerful pulses periodically to average out the effect of unwanted Hamiltonian $H_{SB}$, and achieves the purpose of decoupling with the environment. The BB operation group $G$ is designed to satisfy the dynamic decoupling condition (3). And set

$t_i = \delta_i T$
$(i = 1, 2, \cdots, M; \quad 0 < \delta_1 < \delta_2 < \cdots < \delta_{M-1} < \delta_M < 1)$,

in which the time moments $t_i$ with the pulses of $M$ can be written as

$$t_i = T * \delta_{i-PDD} = T * i / (M+1), \quad (6)$$
$$\delta_{i-PDD} = i / (M+1), \quad i = 1, 2, \cdots, M. \quad (7)$$

### 2.3 Uhrig Dynamical Decoupling (UDD)

The Uhrig dynamical decoupling (UDD) [5] is an optimized dynamical decoupling scheme proposed by Uhrig. It also uses the twin-born tailored powerful pulses to average out the effect of unwanted Hamiltonian $H_{SB}$, and achieves the purpose of decoupling with the environment. The BB operation group $G$ in UDD is also designed to completely cancel errors up to the first order in the Magnus expansion, and should satisfy the dynamic decoupling condition (3). This is just the same with PDD, but the difference from the PDD is that UDD optimizes the time intervals between adjacent pulses, and the $t_i$ with the UDD pulses of $M$ can be expressed as [5]

$$t_i = T * \delta_{i-UDD} \quad (8)$$

$$\delta_{i-UDD} = \sin^2[i * \pi / (2M+2)], \quad i = 1, 2, \cdots, M \quad (9)$$

## 3 Behavior of quantum coherence under Various dynamical decoupling schemes

In order to investigate the performance of suppressing decoherence of these dynamical decoupling schemes, we first derive the trend of the non-diagonal element of density matrix to observe the evolution of the Ξ-type $n$-level atom under dynamical decoupling schemes. Then we choose a Ξ-type six-level atom as a model for system controlled, and derive the trends of the non-diagonal element of density matrix while PDD and UDD are applied respectively, and finally make a comparison and analysis of the results.

### 3.1 Evolution of A Ξ-Type n-Level Atom Under Dynamical Decoupling Schemes

In order to analyze easily the impact of the decoupling pulse sequence, transform the system into the interaction picture, then the interaction Hamiltonian in (2) can be expressed as

$$\tilde{H}_{SB} = \hbar \sum_{i=0}^{n-2} \sum_{ki} \sigma_z^{(i+1,i)} (j_{ki} a_{ki}^+ e^{i\omega_{ki} t} + j_{ki}^* a_{ki} e^{-i\omega_{ki} t}) \quad (10)$$

And the evolution propagator of the whole system can be expressed as

$$\tilde{U}(t_s, t_o) = \exp\{\sum_{i=0}^{n-2} \sum_{ki} \sigma_z^{(i+1,i)} (a_{ki}^+ e^{i\omega_{ki} t_o} \zeta_{ki}(t_o - t_s) - h.c.)\}$$
(11)

where $t_s$ and $t_o$ represent the start and end moments of this evolution process; $\zeta_{ki}(t_o - t_s) = \frac{j(\omega_{ki})}{\omega_{ki}}(1 - e^{i\omega_{ki}(t_o - t_s)})$.

Now consider the situation in the $j^{th}$ ($j=1,2,\ldots,N$) decoupling cycle. According to the full expressions for the BB operators $\{g_k\}$ in (4), we can program a cycle of pulse sequence as $\{g_1, g_1^+, g_2, g_2^+, \cdots, g_{n-2}, g_{n-2}^+, g_{n-1}, g_{n-1}^+\}$. So we get a cycle of the unitary evolution of the atom system under the BB operations as the form of

$$\tilde{U}(t_j^s, t_j^o) = g_{n-1}^+ \tilde{U}(t_j^{(n-1)}, t_j^{(n)}) g_{n-1} g_{n-2}^+ \cdots$$
$$g_2 g_1^+ \tilde{U}(t_j^{(1)}, t_j^{(2)}) g_1 \tilde{U}(t_j^{(0)}, t_j^{(1)}) \quad (12)$$

where $t_j^s$ and $t_j^o$ are the start and end moments in the $j^{th}$ decoupling cycle, and $t_j^s = t_j^{(0)}$, $t_j^o = t_j^{(n)}$.

Let $T_{c,j}$ represent the length of the $j^{th}$ decoupling cycle, then one can get $T_{c,j} = t_j^o - t_j^s$. Let $t_j^{(i)}$ and $\Delta t_j(i)$ ($i = 1,2,\ldots, n$) represent the moment of the pulse $g_i g_{i-1}^+$ and the $i^{th}$ time interval of the adjacent two pluses in the $j^{th}$ decoupling cycle, and one can have $\Delta t_j(i) = t_j^{(i)} - t_j^{(i-1)}$.

Then consider in the total evolution time $T$. After applying periodically the decoupling operations $\{g_k\}$ in (4) in $N$ decoupling cycles, the evolution propagator of the whole system is

$$\tilde{U}(0,T) = \tilde{U}(t_N^s, T)\tilde{U}(t_{N-1}^s, t_{N-1}^o) \cdots \tilde{U}(t_2^s, t_2^o)\tilde{U}(0, t_1^o) \quad (13)$$

As is well known, if the atom is not affected by the environment, the relevant quantity is the qubit coherence $\tilde{\rho}_S^{ij}(T)$ ( $i \neq j$ ), and of course, $\tilde{\rho}_S^{ij}(T) = \tilde{\rho}_S^{ij}(0)$. To observe the suppression effect of the BB decoupling pulse sequences on the atom of the adiabatic decoherence, we are interested in calculating the reduced density matrix of the atom. In this work, we choose and calculate the non-diagonal element $\tilde{\rho}_S^{01}(T)$ for simplicity. The other non-diagonal elements $\tilde{\rho}_S^{ij}(T)$ ( $i \neq j$ ) can also be chosen, but they are not discussed here.

We assume that the atom and environment are initially uncorrelated, i.e.

$$\tilde{\rho}(0) = \tilde{\rho}_S(0) \otimes \tilde{\rho}_B(0) \quad (14)$$

where $\tilde{\rho}_B(0)$ is a kind of thermal equilibrium state at temperature $T'$, i.e.

$$\tilde{\rho}_B(0) = \prod_k (1 - e^{\beta \hbar \omega_k}) * e^{-\beta \hbar \omega_k a_k^+ a_k}$$

where $\beta = 1/(k_B * T')$, $k_B$ is the Boltzmann constant and $T'$ is the temperature of the bath. And one can choose henceforth units such that $\hbar = k_B = 1$ for simplicity. Then combined with (11)-(14), one obtains

$$\tilde{\rho}_S^{01}(T) = \langle 0 | \text{Tr}_B \{\tilde{U}(0,T)\tilde{\rho}(0)\tilde{U}^+(0,T)\} | 1 \rangle$$
$$= \tilde{\rho}_S^{01}(0) * \exp\{-\sum_{m=1}^{n-1} \Gamma_m(\Delta t_j(i), T_{c,j}, T)\} \quad (15)$$

where $i=1,2,\ldots,n$, $j=1,2,\ldots,N$,

$$\Gamma_m(\Delta t_j(i), T_{c,j}, T) = \frac{1}{2} \sum_{ki} |\chi_m(\Delta t_j(i), T_{c,j}, T)|^2 \coth \frac{\omega_{ki}}{2T}$$
(16)

where $m = 1, 2, \cdots, n-1$. $\chi_m(\Delta t_j(i), T_{c,j}, T)$ in (16) can be expressed as

$$\chi_1(\Delta t_j(i), T_{c,j}, T) = -2 * \eta_1(\Delta t_j(i), T_{c,j}, T)$$
$$+ \eta_2(\Delta t_j(i), T_{c,j}, T) + \eta_n(\Delta t_j(i), T_{c,j}, T)$$
(17a)

$$\chi_2(\Delta t_j(i), T_{c,j}, T) = -2 * \eta_2(\Delta t_j(i), T_{c,j}, T)$$
$$+ \eta_1(\Delta t_j(i), T_{c,j}, T) + \eta_{n-1}(\Delta t_j(i), T_{c,j}, T)$$
(17b)

and when $m = 3, 4, \cdots, n-1$, one has

$$\chi_m(\Delta t_j(i), T_{c,j}, T) = -2 * \eta_{n-m+2}(\Delta t_j(i), T_{c,j}, T)$$
$$+ \eta_{n-m+1}(\Delta t_j(i), T_{c,j}, T) \quad , \quad (17c)$$
$$+ \eta_{n-m+3}(\Delta t_j(i), T_{c,j}, T)$$

where

$$\eta_l(\Delta t_j(i), T_{c,j}, T) = \zeta_{ki}(\Delta t_1(l)) * e^{i\omega_{ki}\sum_{i=1}^{l-1}\Delta t_1(i)}$$
$$+ \zeta_{ki}(\Delta t_2(l)) * e^{i\omega_{ki}\sum_{i=1}^{l-1}\Delta t_2(i)} * e^{i\omega_{ki}T_{c,1}} + \cdots$$
$$+ \zeta_{ki}(\Delta t_N(l)) * e^{i\omega_{ki}\sum_{i=1}^{l-1}\Delta t_N(i)} * e^{i\omega_{ki}\sum_{j=1}^{N-1}T_{c,j}}$$

(18)

where $l = 1, 2, \cdots, n$.

Now according to different moments to generate the decoupling pulse sequence in PDD and UDD, we get different $\Delta t_j(i)$ and $T_{c,j}$.

1) when PDD is applied, one has

$$\delta_{0-PDD} = 0, \quad \delta_{(n*N)-PDD} = 1 \quad (19a)$$

$$\Delta t_j(i) = (\delta_{m-PDD} - \delta_{(m-1)-PDD}) * T \quad (19b)$$

where

$$m = (j-1)*n + i, \quad i=1,2,\ldots,n, \quad j=1,2,\ldots,N.$$

$$T_{c,j} = \Delta t_j(1) + \Delta t_j(2) + \cdots + \Delta t_j(n).$$

(19c)

2) when UDD is applied, one has

$$\delta_{0-UDD} = 0, \quad \delta_{(n*N)-UDD} = 1 \quad (20a)$$

$$\Delta t_j(i) = (\delta_{m-UDD} - \delta_{(m-1)-UDD}) * T \quad (20b)$$

where

$$m = (j-1)*3 + i, \quad i=1,2,\ldots,n, \quad j=1,2,\ldots,N.$$

$$T_{c,j} = \Delta t_j(1) + \Delta t_j(2) + \cdots + \Delta t_j(n).$$

(20c)

We know that the spectral density $I(\omega) \to 0$ when the frequency $\omega$ is greater than the finite cut-off frequency of each mode of the environment $\omega_c$, i.e., $\omega > \omega_c$ [3]. When the bath is the generally considered Ohmic bath, we can have the spectral density for each mode of the bath to be $I(\omega) = \frac{\alpha}{4}\omega^r e^{-\omega/\omega_c}$, where $\alpha$ measures the strength of the system-environment interaction and $r = 1$ [3],[11]-[13]. The following transformation [3], [14] can be taken in the continuum limit of the bath mode,

$$\sum_{ki} \to \int_0^{\omega_c} d\omega_{ki} I(\omega_{ki}) \frac{1}{|j(\omega_{ki})|^2}.$$

Then (16) changes into

$$\Gamma_m(\Delta t_j(i), T_{c,j}, T) = \frac{1}{2}\int_0^{\omega_c} d\omega_{ki} I(\omega_{ki}) \coth\frac{\omega_{ki}}{2T} \left|\frac{\chi_m(\Delta t_j(i), T_{c,j}, T)}{j(\omega_{ki})}\right|^2$$
(21)

Set

$$P(T) = \exp\{-\sum_{m=1}^{n-1}\Gamma_m(\Delta t_j(i), T_{c,j}, T)\} \quad (22)$$

where $\Gamma_m(\Delta t_j(i), T_{c,j}, T)$ is derived from (21).

So (15) is changed into

$$\tilde{\rho}_S^{01}(T) = \tilde{\rho}_S^{01}(0) * \exp\{-\sum_{m=1}^{n-1}\Gamma_m(\Delta t_j(i), T_{c,j}, T)\} = \tilde{\rho}_S^{01}(0) * P(T)$$
(23)

From (23) one can see that $P(T)$ represents the attenuation ratio of the non-diagonal element $\tilde{\rho}_S^{01}(0)$ at the moment $T$ under dynamical decoupling schemes and none control schemes, respectively. It is easy to see that when $P(T)$ is closer to 1, the better suppression of decoherence will be got. Thus we can observe the curves of the function $P(T)$ to understand the loss of the quantum coherence when PDD and UDD are applied, respectively. And in part B, the numerical simulations will be carried out in a six-level atom in Ξ-configuration.

## 4 Illustration Example and Result Analysis

Consider a six-level atom in Ξ-configuration as shown in Fig. 1. The six levels are labeled as $|0\rangle, |1\rangle, |2\rangle, |3\rangle, |4\rangle$ and $|5\rangle$, respectively, and their energies are $E_0$, $E_1$, $E_2$, $E_3$, $E_4$ and $E_5$. The level $|k\rangle$ is coupled to the level $|k+1\rangle$ via the fields of resonance frequencies $\omega_{k+1\,k}$, ($k = 0,1,\cdots,4$), respectively. We place the atom under the driving of five fields with frequencies of $\omega_{j\,j-1} = (E_j - E_{j-1})/\hbar$, ($1 \le j \le 5$), respectively. Similarly we define frequencies $\omega_{j\,j-2}(j-2 \ge 0),\cdots,\omega_{j\,j-n}(j-5 \ge 0)$, ($j = 2,3,\cdots,5$).

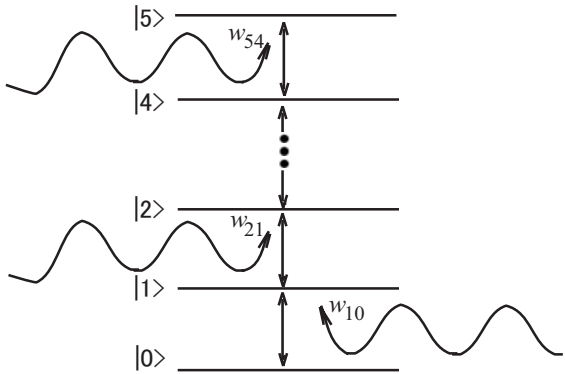

Fig. 1 A six-level atom in Ξ-configuration under five driving fields with frequencies $\omega_{10}$, $\omega_{21}$, $\omega_{32}$, $\omega_{43}$ and $\omega_{54}$.

According to part A, here one can have $n = 6$. Thus from (22) one can obtain

$$P(T) = \exp\{-\sum_{m=1}^{5} \Gamma_m(\Delta t_j(i), T_{c,j}, T)\} \qquad (24)$$

Now insert (19) and (20) to (24), respectively, the numerically simulation experiments of the function $P(T)$ under PDD and UDD with the same other simulation parameters are done. And the results are shown in Fig. 2.

In Fig. 2, the dotted line represents the evolution of $P(T)$ while PDD is applied, the solid line represents the evolution of $P(T)$ while UDD is applied. The correlated parameters are $\alpha = 0.25$, $T' = 150\text{K}$, $N = 50$, and $\omega_c = 100\text{Hz}$.

From Fig. 2 one can see that under the same parameters, UDD gets better performance of suppressing decoherence than PDD. The results reveal that UDD is optimized. It means that UDD enhances the possible storage time by up to several times below a certain threshold of the loss of the quantum coherence. And the number of pulses required to obtain a certain prolongation of the storage time with UDD can be much smaller than the number with the standard scheme (PDD).

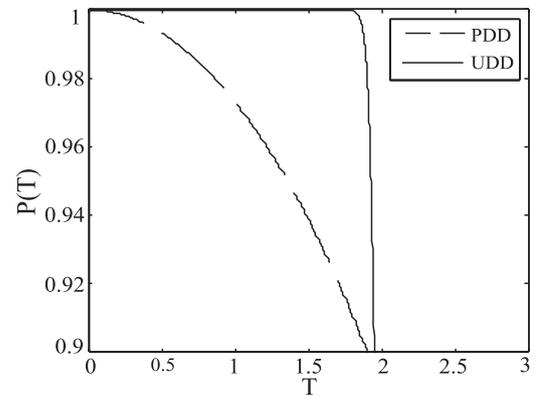

Fig. 2 The curves of $P(T)$ while applying PDD and UDD

## 4 Conclusion

We have investigated the suppression of decoherence of an arbitrary $n$-level atom in Ξ-configuration by means of the dynamic decoupling methods in this paper. Through the comparison simulation experiments, the results show that UDD offers superior performance to PDD over a range of experimentally relevant parameters, such as the temperature of the bath $T'$, the cycle number $N$, the cut-off frequency $\omega_c$ and the strength of the system-environment interaction $\alpha$. It means that the optimized scheme (UDD) can obtain more storage time than the standard one (PDD) below a certain threshold of the loss of the quantum coherence when the same number of pulses are applied. This is very helpful to practical implementations of quantum computation. And by selecting the non-diagonal

ISBN: 978-960-474-372-8    33

element of density matrix as a reference indicator, the comparison of the results also reveal that UDD enhances the possible storage time by up to several times below a certain threshold of the loss of the quantum coherence. Alternatively, the number of pulses required to obtain a certain prolongation of the storage time can be much smaller than the number in the standard scheme (PDD).

Compared to the standard equidistant pulse sequences, it is easy to see that the optimized dynamical decoupling scheme (UDD) is just finding a good way to optimize the time intervals between the adjacent BB pulses. So it will be interesting to research how to optimize the time intervals between the adjacent BB pulses. We have also completed the researches on suppression of general decoherence in arbitrary n-level atom in Ξ-configuration under bang-bang control [15] and an optimized dynamical decoupling strategy to suppress decoherence [16]. The more study on quantum system control can be find in [17].

# 5 Acknowledgments

This work was supported in part by the National Key Basic Research Program under Grant No. 2011CBA00200